\documentclass[12pt,preprint]{aastex}

\shorttitle{Protostellar Collapse Triggered by a Small HII Region}

\shortauthors{ Kang \& Kerton}

\begin{document}
\title{IRAS 01202+6133 : A Possible Case of Protostellar Collapse\\
Triggered by a Small HII Region}

\author{Sung-Ju Kang and C. R. Kerton}

\affil{Department of Physics \& Astronomy, Iowa State University,
  Ames, IA 50011, USA}
\email{sjkang@iastate.edu, kerton@iastate.edu}

\begin{abstract}

We present an analysis of HCO$^{+}$ ($J=3\rightarrow2$) and H$^{13}$CO$^{+}$ ($J=3\rightarrow2$)
observations of the massive (M$\sim20$ M$_\sun$) submm/IR source IRAS
01202+6133 located on the periphery of the \ion{H}{2} region KR~120
(Sh 2-187).  The HCO$^{+}$ line profile has a classic blue-asymmetric
shape with the optically thin H$^{13}$CO$^{+}$ line peaking at the
position expected if the HCO$^{+}$ line arises from a combination of
self-absorption and infall motion. We have modified existing analytic
radiative transfer models to allow for the fitting of submm/mm line
profiles that have both self-absorption features and optically thin
wings and applied these models to our HCO$^{+}$ spectrum of
IRAS~01202+6133. We conclude that it is a young Class I YSO with a
substantial envelope undergoing slow infall and having some outflow
motions. The young age of the \ion{H}{2} region rules out
a ``collect and collapse'' scenario. While we cannot eliminate the possibility that
IRAS~01202+6133 formed spontaneously at its current location,
considering its early evolutionary state and its proximity to the
\ion{H}{2} region we think that the formation
of IRAS~01202+6133 was triggered by the expansion of KR 120 (Sh 2-187).
 
\end{abstract}


\section{Introduction} \label{sec:intro}

\begin{figure}
\plotone{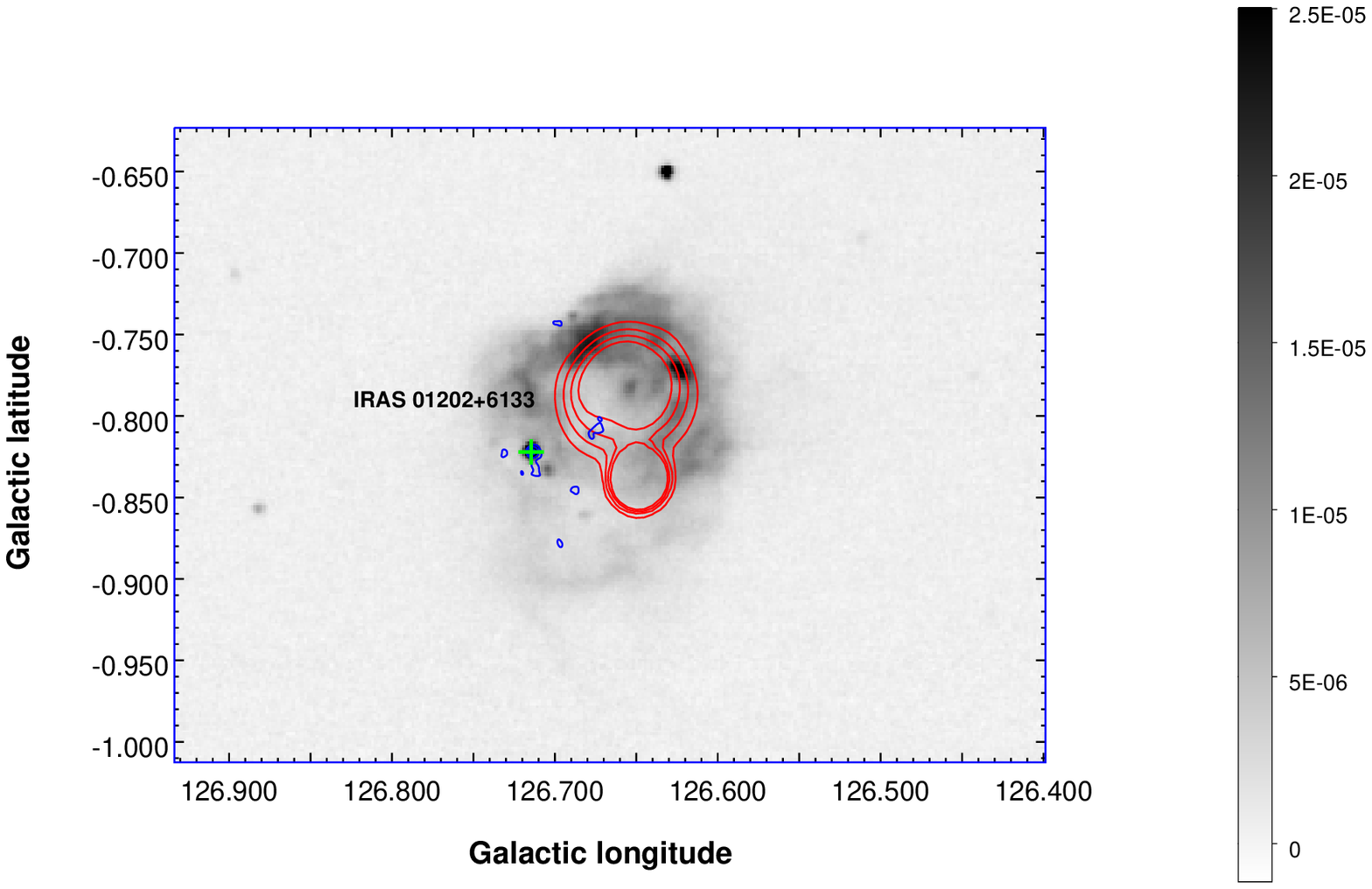}
\caption{The \ion{H}{2} Region KR 120 (Sh2-187). This \emph{Midcourse Space Experiment} \citep[\emph{MSX};][]{pri01} 8.3~$\mu$m image traces emission from the
  photodissociation region within the molecular cloud surrounding KR 120. The red contours correspond to
  Canadian Galactic Plane Survey \citep[CGPS;][]{tay03} 1420 MHz
  continuum emission at $T_B = $ 8, 10, 12 and 14 K. The elongated radio source seen at the bottom of the main
    \ion{H}{2} region is extragalactic.  The blue contours represent SCUBA 850 $\mu$m continuum with levels of 
  $1.2\times10^{-3},\ 2\times10^{-3},\ 3\times10^{-3},\
  6\times10^{-3}$ Jy beam$^{-1}$. Note the location of the
  Class I YSO IRAS 01202+6133 on the periphery of the \ion{H}{2}
  region.  \label{fig:kr120} }
\end{figure}

The molecular gas surrounding an \ion{H}{2} region is thought to be a place
where star formation can be induced. Such triggered star formation can
arise from the overpressurization of existing density enhancements
\citep{tho04} or through the collapse of a swept up layers of material
\citep{elm98,deh03}. This paper presents an analysis of submm
spectroscopic observations of the Class I Young Stellar Object (YSO)
IRAS~01202+6133 which is located on the periphery of the relatively
nearby ($d=1.44\pm0.26$~kpc) \ion{H}{2} region KR 120 (Sh 2-187;
\citealt{arv11,kal80,sha59}). As shown in Figure~\ref{fig:kr120} the proximity
alone of this massive ($M=21\pm9$~M$_\sun$), luminous ($L\sim
5600$~L$_\sun$) YSO to the \ion{H}{2} region makes it a strong
candidate for an example of triggered star formation. The submm
spectra presented here strengthen this scenario as they clearly show
infall is occurring as would be expected for a very young YSO. We
believe that IRAS~01202+6133 is one of the clearest examples to date of an
\ion{H}{2} region triggering the formation of another massive star.

Our observations are described in \S~\ref{sec:obs}. In
\S~\ref{sec:models} we review the analytic models of \citet{mye96} and
\citet{dev05} and develop new varieties of these models that allow for
both infall and outflow motions. Physical properties of the YSO
derived using these models are presented and discussed in
\S~\ref{sec:results} followed by our conclusions in \S~\ref{sec:conc}.

\section{Observations}\label{sec:obs}

\begin{figure}
\plotone{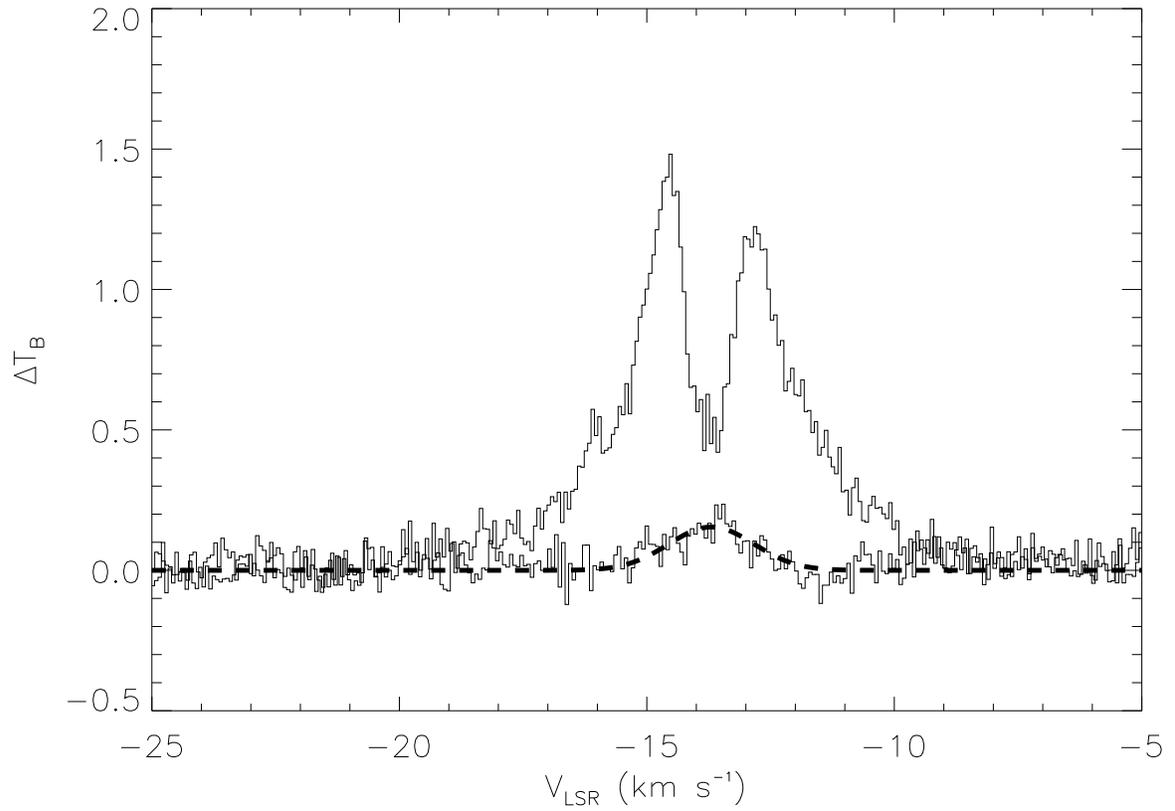}
\caption{HCO$^{+}$ spectra of IRAS~01202+6133. The HCO$^{+}$ ($J = 3\rightarrow 2$; 267.6 GHz) 
line has a classic blue asymmetric line profile. The optically thin
H$^{13}$CO$^+$ ($J = 3\rightarrow 2$; 260.3 GHz) line (with Gaussian fit
shown) peaks at the minimum between two peaks of HCO$^{+}$ line
confirming that blue asymmetric line shape comes from the infall motion. \label{fig:spec}}
\end{figure}

In order to explore the kinematics of IRAS~01202+6133 we obtained
observations of the commonly used \citep[e.g.,][]{sun09} optically
thick infall tracer molecule HCO$^+$ ($J = 3\rightarrow 2$; 267.6 GHz)
and its optically thin isotopologue H$^{13}$CO$^+$ ($J = 3\rightarrow
2$; 260.3 GHz) at the James Clerk Maxwell Telescope (JCMT) in the 06B
semester. Single pointing observations (beamsize $\sim 20\arcsec$
FWHM) were made using the A3 receiver and ACSIS backend with a
frequency resolution of 61 kHz (0.068 km s$^{-1}$) and an integration
time of $\sim 40$ minutes for each line. The reduced spectra are shown in
Figure ~\ref{fig:spec}. Both have a noise level of $\sim 0.1$~K
resulting in a peak S/N$\sim 30$ for the HCO$^+$ spectrum and $\sim 3$
for the weaker H$^{13}$CO$^+$ spectrum.

If there was no infall motion associated with IRAS~01202+6133 we
would expect the HCO$^+$ spectrum to be a single self-absorbed profile
with symmetric red and blue peaks around a local minimum at some
systematic velocity. Following \citet{mye96}, in the case of where
there is infall there will be density and excitation temperature ($T_{ex}$) differences
between the central regions and the outer parts of the molecular
core. Blueshifted emission from the rear of the core is not strongly
absorbed because it passes through the central (high $T_{ex}$) region
then through a highly Doppler shifted front portion of the core. In
contrast the redshifted emission from the front of the core is
absorbed by surrounding gas that is at both a comparable velocity and
similar (low) $T_{ex}$.   The overall effect of the infall motion is
to create a ``blue-asymmetric'' line profile where the self-absorption
line is asymmetric with a stronger blue peak. Such a line profile
could also form from multiple unrelated emission components being
aligned along the line of sight or from rotational motion, but in both
cases we would expect to see a similar profile in the optically thin
line. The fact that our  H$^{13}$CO$^{+}$ spectrum peaks up at the same
velocity as the minimum in the HCO$^{+}$ spectrum makes us confident
that we are looking at a single self-absorbed line profile with the
asymmetry arising from infall motions.

\section{Analytic Radiative Transfer Modeling}\label{sec:models}

\citet{mye96} and \citet{dev05} developed basic one-dimensional (1D)
analytic radiative transfer models to derive kinematic data from mm/submm spectra of
star-forming molecular cores. In this section we first review these
models and then present modified versions that better model cases where
there is simultaneous infall and outflow motions.

All of the models are based on the general solution to the 1D radiative
transfer equation for a homogeneous medium with total optical depth
$\tau_0$, 
\begin{equation}\label{eqn:gsol}
T_B = T_ie^{-\tau_0}+\int_0^{\tau_0}J(T)e^{-\tau}d\tau
\end{equation}
where $T_B$ is the brightness temperature of the outgoing radiation,
$T_i$ is the brightness temperature of any incoming radiation, $J(T)
\equiv T_{0}$/[exp$(T_{0}/T)-1]$ and $T_{0} \equiv h \nu / k$, where $\nu$ is
the frequency of the transition and $T$ is the temperature.

If $J(T)$ is a linear function of $\tau$ then equation~\ref{eqn:gsol}
can be integrated to obtain, 
\begin{equation}\label{eqn:lsol}
T_B = T_ie^{-\tau_0}+(J_2-J_1)\frac{1-e^{-\tau_0}}{\tau_0}+J_1-J_2e^{-\tau_0}
\end{equation}
where $J_1$ and $J_2$ are the values of $J(T)$ at the two ends of the
path with optical depth $\tau_0$. All of the models presented in this
section are based on equation~\ref{eqn:lsol} with variations due
to how $J(T)$ varies with $\tau$ in detail.

\subsection{Two-layer model}\label{sec:twolayer}

The two-layer model of \citet{mye96} has ``front'' and ``rear''
constant $T_{ex}$ layers moving toward each other each at an infall
velocity, $v_{in}$ (see Figure 1 of \citealt{dev05}).

The line brightness temperature of the two-layer model is 
\begin{equation} \label{eqn:2Ldt}
\Delta T_B =  J(T_f)[1-e^{-\tau_f}]+J(T_r)[1-e^{-\tau_r}]e^{-\tau_f} - J(T_b)[1-e^{-\tau_f-\tau_r}] 
\end{equation}
where $T_f$ and $T_r$ are the front and rear excitation temperatures
and $T_b$ is the background temperature. The optical depths of each
layer, $\tau_f$ and $\tau_r$ are given by
\begin{mathletters}
\begin{eqnarray}
\tau_f  & = & \tau_0\ \mbox{exp} \ [-(v-v_{LSR}-v_{in})^{2}/2\sigma^2] , \label{eqn:2Ltauf} \\
\tau_r  & = & \tau_0\ \mbox{exp} \ [-(v-v_{LSR}+v_{in})^{2}/2\sigma^2] \label{eqn:2Ltaur}
\end{eqnarray}
\end{mathletters}
where $\tau_{0}$ is the line center optical depth, $v_{LSR}$ is a average
line-of-sight velocity of the system, and $\sigma$ is the velocity
dispersion of the observed molecule.

For a given choice of $T_b$ (usually $T_b = 2.7$~K) the two-layer
model has six adjustable parameters: $T_f$, $T_r$, $\tau_0$,
$v_{LSR}$, $v_{in}$, and $\sigma$. By making some assumptions about the
expected density and temperature structure in an infalling core
\citet{mye96} was able to reduce the number of parameters in the two-layer
model to five, replacing $T_r$ and $T_f$ by a kinetic temperature,
$T_k$. This version of the model also used the non-thermal velocity
dispersion as an input, $\sigma_{NT} \equiv
(\sigma^{2}-kT_{k}/m_{obs})^{1/2}$ where $m_{obs}$ is the mass of the
observed molecule. Here we present the formulas 
used to calculate the model parameters and refer the reader to
\citet{mye96} for their derivation.

The excitation temperature of each layer is determined using the following equations, 
\begin{mathletters}
\begin{eqnarray}
\frac{T_f}{T_k} & = & \frac{T_b+(4T_0/\beta)\langle n_f  \rangle/n_{max}}{T_k+(4T_0/\beta)\langle  n_f\rangle /n_{max}} , \label{eqn:2Ltf} \\
\frac{T_r}{T_k} & = & \frac{T_b+(4T_0/\beta)\langle n_r \rangle/n_{max}}{T_k+(4T_0/\beta)\langle n_r\rangle /n_{max}}, \label{eqn:2Ltr}
\end{eqnarray}
\end{mathletters}
where $\beta =[1-\mbox{exp}(-\tau_{0})]/\tau_{0}$, and $\langle n_f
\rangle/n_{max}$ and $\langle n_{r}\rangle/n_{max}$ are given by, 
\begin{mathletters}
\begin{eqnarray}
\frac{\langle n_f  \rangle}{n_{max}} & = &
(1-e^{-\tau_0})^{-1}\left[\frac{6}{\tau_0^3}-e^{-\tau_0}(1+\frac{3}{\tau_0}+\frac{6}{\tau_0^2}+\frac{6}{\tau_0^3})\right
] , \label{eqn:2Lnf} \\ 
\frac{\langle n_r  \rangle}{n_{max}} & = & (1-e^{-\tau_0})^{-1}[1-\frac{3}{\tau_0}+\frac{6}{\tau_0^2}-\frac{6}{\tau_0^3}(1-e^{-\tau_0})]. \label{eqn:2Lnr}
\end{eqnarray}
\end{mathletters}

\subsection{Hill model}\label{sec:hill}

In the ``hill'' model of \citet{dev05} $J(T)$ increases
linearly from the $T_0$ to $T_p$ over a front optical depth
($\tau_f$), and then decreases from $T_p$ to $T_0$ over a rear optical depth
($\tau_r$). The shape of a $J(T)$ plot against $\tau$ gives the model
its name (see Figure~1 of \citealt{dev05}). As with the two-layer
model the front and rear sections each have a line center optical
depth ($\tau_0$), are moving toward each other with an infall velocity
$v_{in}$, and have a velocity dispersion $\sigma$.

The line brightness temperature of the hill model\footnote{The  equation in \citet{dev05} is typeset incorrectly in the  Astrophysical Journal. The correct version is found in the arXiv  version \citep{dev04}} is given by, 
\begin{equation}\label{eqn:hilldt2}
\Delta T_{B} =
\left[J(T_{p})-J(T_{0})\right]\left[\frac{1-e^{-\tau_f}}{\tau_f}
  -\frac{e^{-\tau_f}(1-e^{-\tau_r})}{\tau_r}\right] + \left[J(T_0)-J(T_b)\right]\left(1-e^{-\tau_f-\tau_r}\right).
\end{equation} 
For a given choice of $T_b$, the hill model has six adjustable
parameters: $T_0$, $T_p$, $\tau_0$, $v_{LSR}$, $v_{in}$, and $\sigma$. 
Equations~\ref{eqn:2Ltauf} and \ref{eqn:2Ltaur} are used to determine $\tau_f$ and $\tau_r$.

\subsection{Analytical models with outflow} \label{sec:outflows}

\begin{figure}
\plottwo{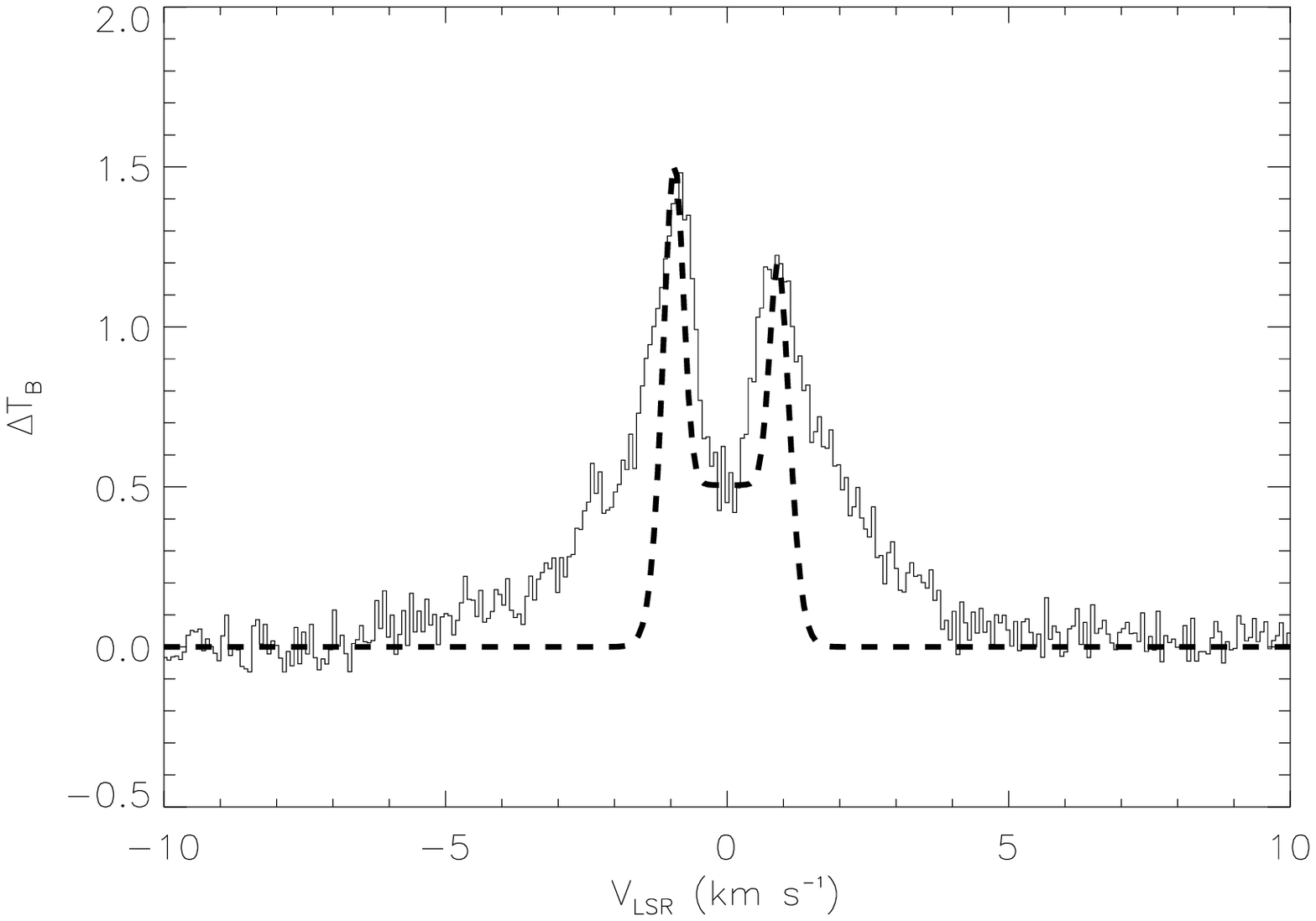}{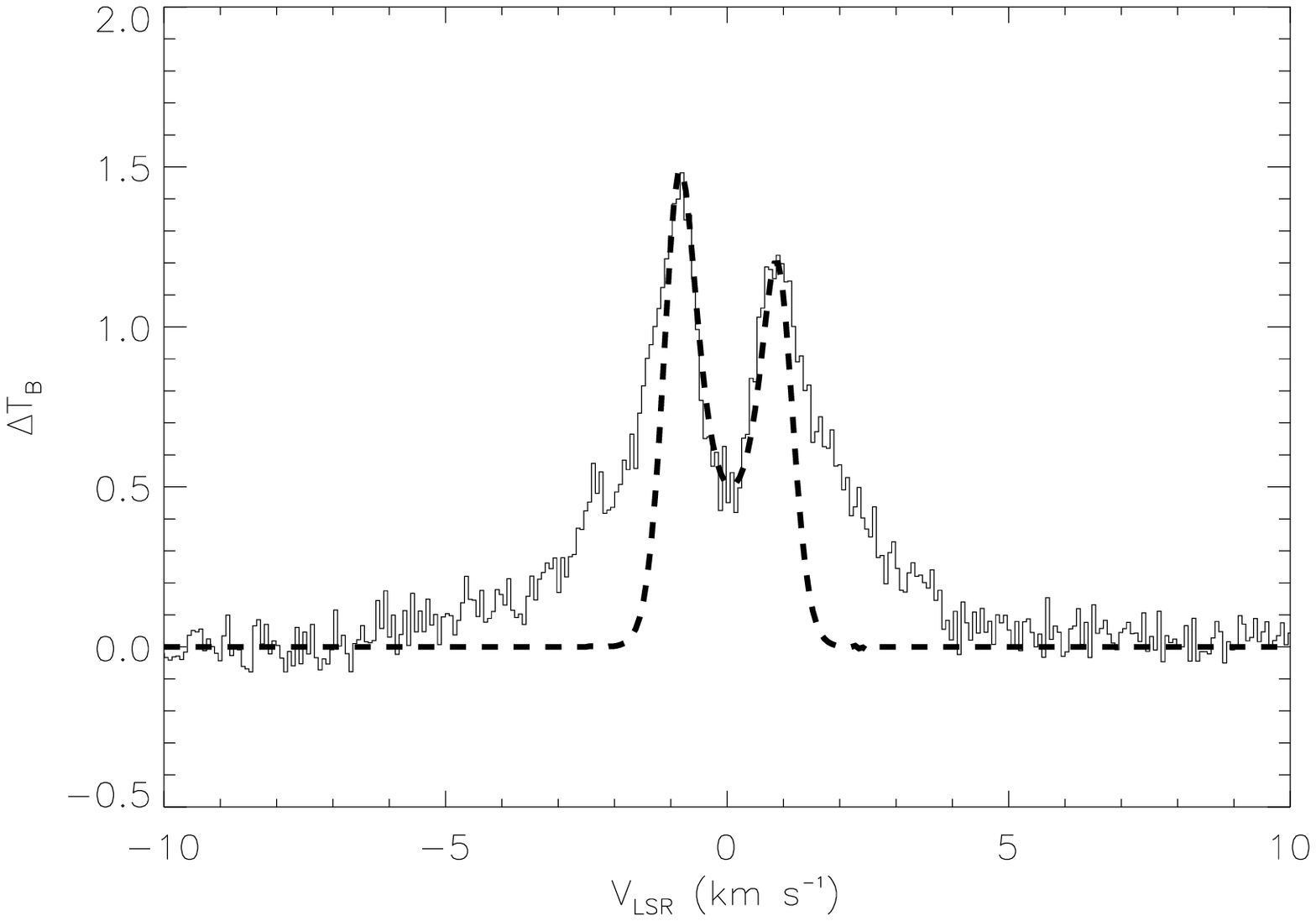}
\caption{Basic model results. The figure shows how the basic Two-Layer
  model (left) and Hill model
(right) are unable to fit the extended wings seen in the
HCO$^{+}$ spectrum. In these plots the systematic velocity of the
cloud ($V_{LSR}=-13.7$ km s$^{-1}$) has been removed. \label{fig:noflow}}
\end{figure}

\begin{figure}
\plottwo{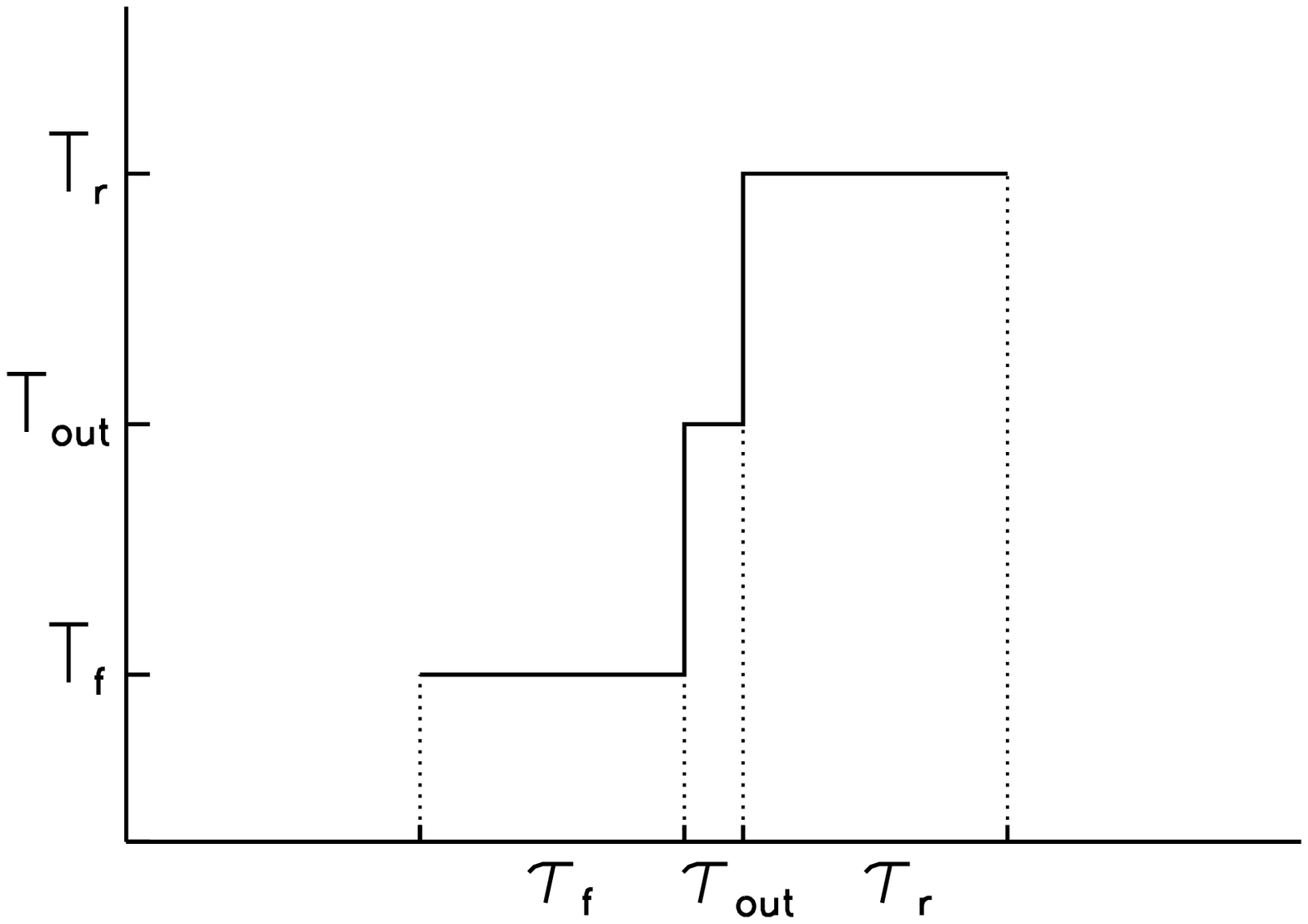}{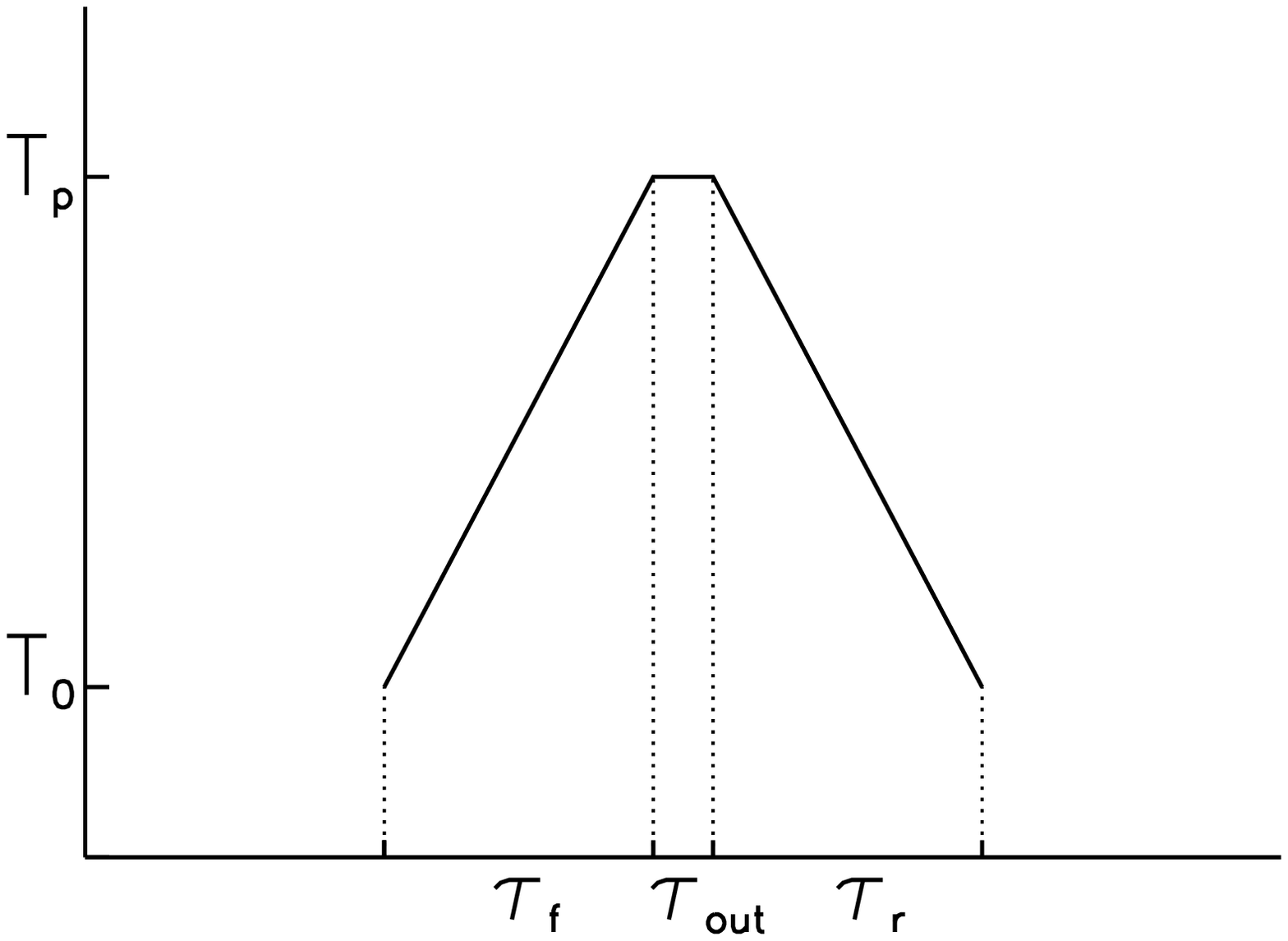}
\caption{Modified radiative transfer models. In order to fit the
  extended wings seen in the HCO$^{+}$ spectrum we added a central low
  optical depth  ``outflow'' region to the exisiting two-layer and hill models. The diagrams
  schematically show the run of temperature with optical depth in the
  2L-O model (left) and Hill-O model (right). \label{fig:models}}
\end{figure}

While the two-layer and the hill models are able to reproduce the central portion of the observed
blue-asymmetric line profile they are not able to model the extended
wings of the line profile observed beyond $\sim\pm2$~km~s$^{-1}$ (see
Figure~\ref{fig:noflow}). In order to correctly model the entire line
we follow the approach used in the \citet{mye96} analysis of L~1527
and incorporate a central outflow region into both the two-layer and hill models. The resulting
variation of $T_{ex}$ with $\tau$ is shown schematically in
Figure~\ref{fig:models} (cf. Figure 1 of \citealt{dev05}).

\subsubsection{Two-layer model with central outflow (2L-O model)}\label{sec:2lo}

With the addition of a central outflow region to the two-layer model
the line brightness temperature (equation~\ref{eqn:2Ldt}) becomes,
\begin{eqnarray} \label{eqn:2Lodt}
\Delta T_B & = &
J(T_f)[1-e^{-\tau_f}]+J(T_{out})[1-e^{-\tau_{out}}]e^{-\tau_f} +J(T_r)[1-e^{-\tau_r}]e^{-\tau_f-\tau_{out}} \\ \nonumber
&  & -J(T_b)[1-e^{-\tau_f-\tau_{out}-\tau_r}]
\end{eqnarray}
where $\tau_{out}$ and $T_{out}$ are the optical depth and
excitation temperature of the outflow. 

The only additional parameters needed in the 2L-O model are
$\tau_{0,out}$ and $\sigma_{out}$ the line center optical depth and
the velocity dispersion of the outflow region. Both are used to
calculate $\tau_{out}$ using,
\begin{equation}\label{eqn:tauo}
\tau_{out} = \tau_{0,out}\exp\left[-(v-v_{LSR})^2/2\sigma_{out}^2\right].
\end{equation}
In our model we set $T_{out}$ equal to the average of $T_r$ and $T_f$ calculated using
$\tau_{out,0}$ in equations~\ref{eqn:2Ltf}, \ref{eqn:2Ltr},
\ref{eqn:2Lnf}, and \ref{eqn:2Lnr} as appropriate.
 
As a test of our version of the 2L-O model we were able to exactly match the
model H$_{2}$CO ($J=2\rightarrow1$) profile of L~1527 shown in
Figure~3 of \citet{mye96} using the outflow and model parameters
given in the figure caption.

\subsubsection{Hill model with central outflow (Hill-O model)}\label{sec:hillo}

The brightness temperature of the hill model with a central outflow is
\begin{equation}\label{eqn:hillodt}
\Delta T_{B} = (J_p-J_0)\left[\frac{1-e^{-\tau_f}}{\tau_f}-\frac{(1-e^{-\tau_r})e^{-\tau_f-\tau_{out}}}{\tau_r}\right]
 + (J_0-J_b)(1-e^{-\tau_f-\tau_{out}-\tau_r})
\end{equation}
where $\tau_{out}$ is the optical depth of outflow. Since we set the
excitation temperature of the outflow equal to $T_p$ (see
Figure~\ref{fig:models}) the only additional parameters are
$\tau_{0,out}$ and $\sigma_{out}$. As in the 2L-O model, $\tau_{out}$
is calculated using equation~\ref{eqn:tauo}.

\section{Results \& Discussion}\label{sec:results}

\subsection{YSO Properties}

\begin{figure}
\plotone{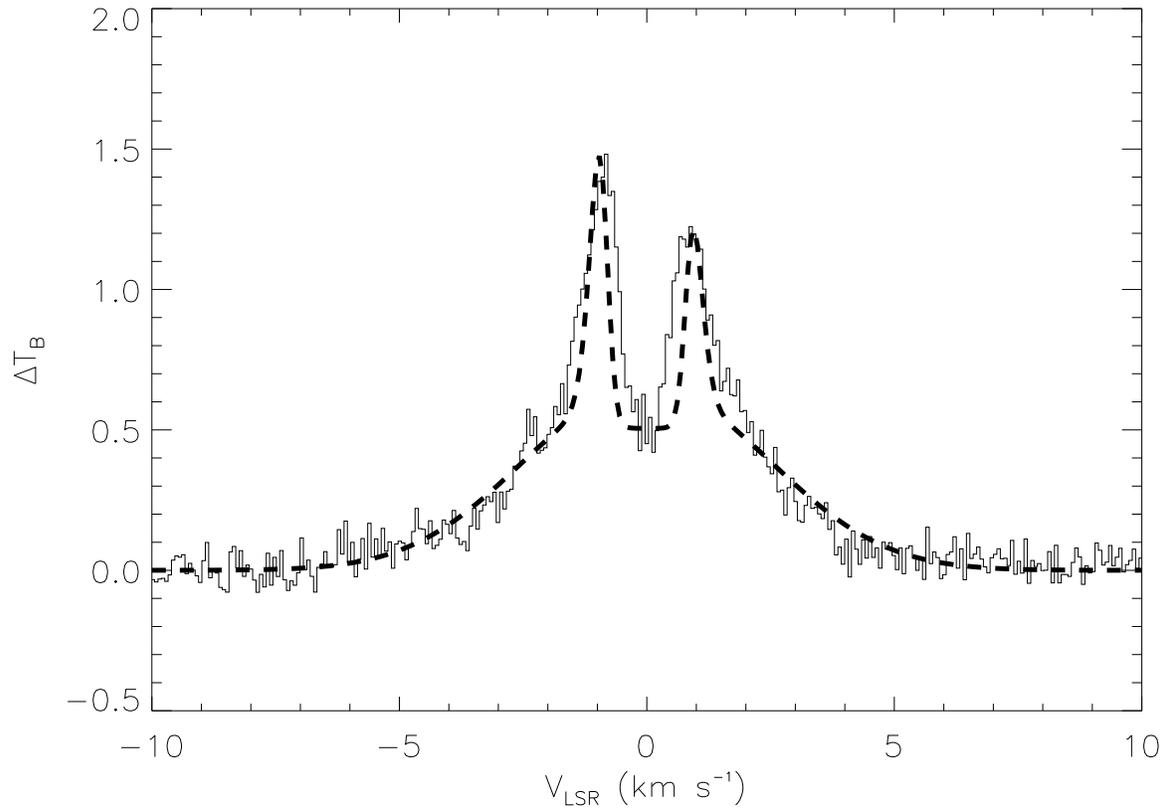}
\caption{Best fit 2L-O model. The model is able to fit the extended
  wings of the HCO$^{+}$ line, but as expected for a two-layer model has some trouble
  matching the line widths and the trough structure. The best fit
  parameters are given in Table~\ref{tbl:best2lo} and the systematic
  velocity of the YSO has been removed from this plot. \label{fig:2lobest}}
\end{figure}

\begin{figure}
\plotone{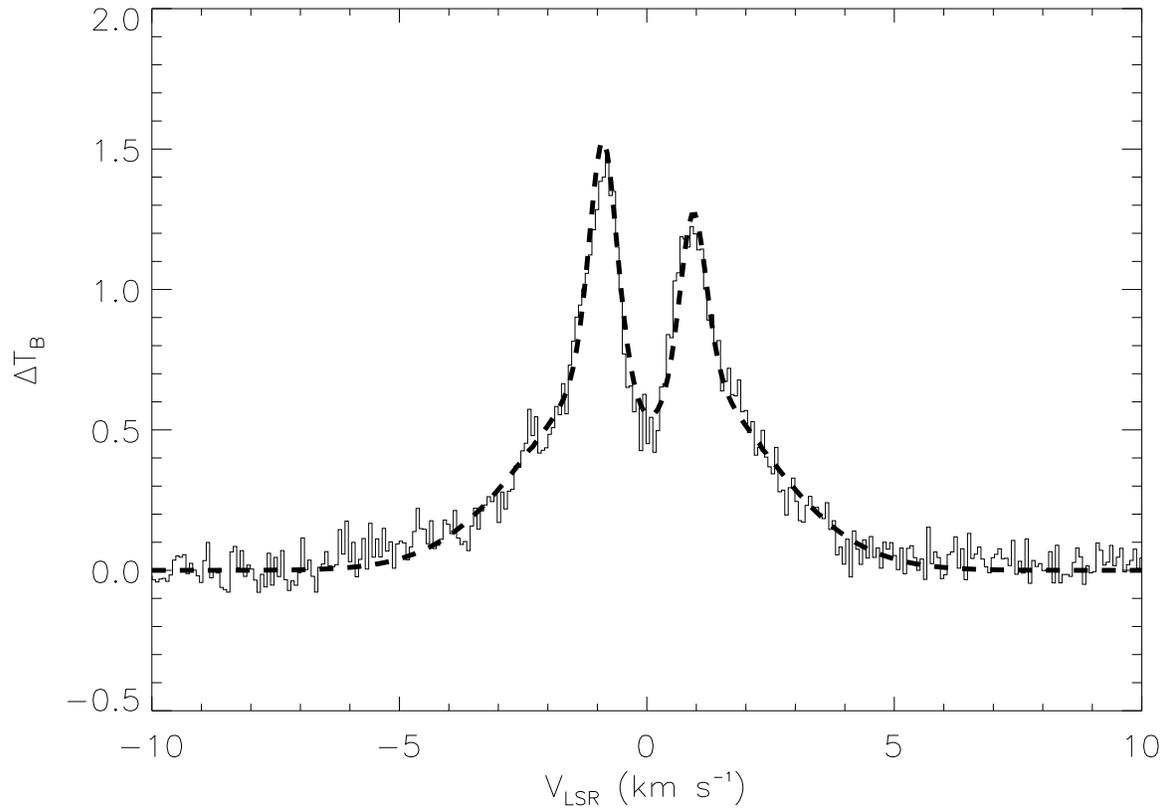}
\caption{Best fit Hill-O model. The model is able to fit the extended
  wings of the HCO$^{+}$ line and has a better fit to the central
  portions of the line profile compared to the 2L-O model. The best fit
  parameters are given in Table~\ref{tbl:besthillo} and the systematic
  velocity of the YSO has been removed from this plot. \label{fig:hillobest}}
\end{figure}

We applied the 2L-O and the Hill-O models to our observed HCO$^{+}$
spectrum (see Figures \ref{fig:2lobest} and \ref{fig:hillobest}
respectively). The best-fit parameters are given in
Tables~\ref{tbl:best2lo} and \ref{tbl:besthillo}. 
Uncertainties were derived using a Monte
Carlo method; an initial best fit of the model was
obtained, then random noise at the same level as the observed line
profile was added to the best fit and the resulting simulated line profile was
then fit generating a new set of parameters. The standard deviation of
spectral line parameters after 1000 repeats was combined with the mean
fit uncertainties to produce the final error estimates.

As noted by \citet{mye96} the two-layer model tends to output profiles
that have narrower peaks and flatter central troughs than
observations, and both of these characteristics can be seen in the
2L-O best-fit model shown in Figure~\ref{fig:2lobest}. In contrast the
Hill-O model does a much better job in fitting the observed peak widths
and the structure of the central absorption trough.

\citet{dev05} showed that Hill style models were able to obtain
accurate estimates of infall velocities (rms error 0.01 km~s$^{-1}$)
from two-peak profiles in contrast with two-layer models, which tended
to underestimate the infall velocity by factors of $\sim 2$.
Consistent with this finding, we see that the
best-fit 2L-O $V_{in}$ is a factor of 3.5 lower than the Hill-O value.
While we adopt the numerical results of the Hill-O model for
IRAS~01202+6133 we note that, regardless of the exact numeric values,
both the 2L-O and Hill-O results are consistent with an object with a
substantial envelope ($\tau_0 \sim 10$) undergoing slow infall ($V_{in} \ll \sigma$).

\subsection{Formation Scenarios}

There are two basic scenarios for the formation of IRAS~01202+6133,
either its formation was triggered by the action of the expanding
\ion{H}{2} region or it formed spontaneously and its proximity to the
\ion{H}{2} region is coincidental. 

To gauge the likelihood of the first scenario we can first look at the
various time scales associated with the YSO-\ion{H}{2} region
system. The canonical duration of the Class I stage of YSO evolution
is of order $10^5$ years \citep{war02}. The fact that IRAS~01202+6133
still has a substantial envelope and exhibits slow infall suggests that it
is still a fairly young Class I YSO. The presence of outflow motions does
not constrain the age as they are known to be associated with even
the very earliest stages of YSO evolution \citep{and93}. Assuming that any Class 0 stage
would be very rapid ($\sim 10^4$ years), we conclude that the age of
the YSO is $\tau_{YSO} \leq 10^5$ years.  KR~120 is a fairly young
\ion{H}{2} region, \citet{jon92} estimate an age $\tau_{HII} \approx
1-2\times10^5$ years based on kinematic and photodissociation region models for the
region. So it appears that a necessary condition for triggered
formation,  that $\tau_{YSO} < \tau_{HII}$, is just satisfied in this case.

As mentioned in \S~\ref{sec:intro}, triggered star formation around
\ion{H}{2} regions can arise from the overpressurization of existing
cores or through the collapse of a swept up layers of material. To
investigate the latter ``collect and
collapse'' scenario we use the equation from \citet{whi94} giving the
timescale at which fragmenation and collapse will occur in a swept-up
shell around an expanding \ion{H}{2} region: 
\begin{equation}\label{eqn:frag}
t_{fragment} \thickapprox 1.56\ \mbox{Myr}\ \ \left[\frac{a_{s}}{0.2 \
    \mbox{km s}^{-1}}\right]^{7/11} \left[\frac{Q_{0}}{10^{49} \
    \mbox{ph s}^{-1}}\right]^{-1/11}\left[\frac{n_{0}}{1000 \
    \mbox{cm}^{-3}}\right]^{-5/11} ,
\end{equation}
where $a_{s}$ is sound speed of the molecular cloud, $Q_{0}$ is the \ion{H}{1} ionizing flux, and $n_{0}$ is the density of
the molecular cloud. The exciting star of KR~120 is B0~V \citep{arv11}
corresponding to $Q_{0}$=10$^{47.4}$ ph s$^{-1}$ using the
\citet{cro05} calibration. For $n_{0}$ we use the value
$n_{0}=1000$ cm$^{-3}$ from \citet{jon92}.
A fragmetation collapse timescale is approximately between
2 Myr with sound speed $a_{s}=0.2$ km s$^{-1}$ (very cold molecular
clouds) and up to a 6 Myr upper limit with with $a_{s}=1$ km s$^{-1}$ (warm molecular/neutral
clouds).
Since this fragmentation timescale is 10 times longer than
$\tau_{HII}$ we can eliminate the ``collect and collapse'' triggered
star formation scenario.

\begin{figure}
\plotone{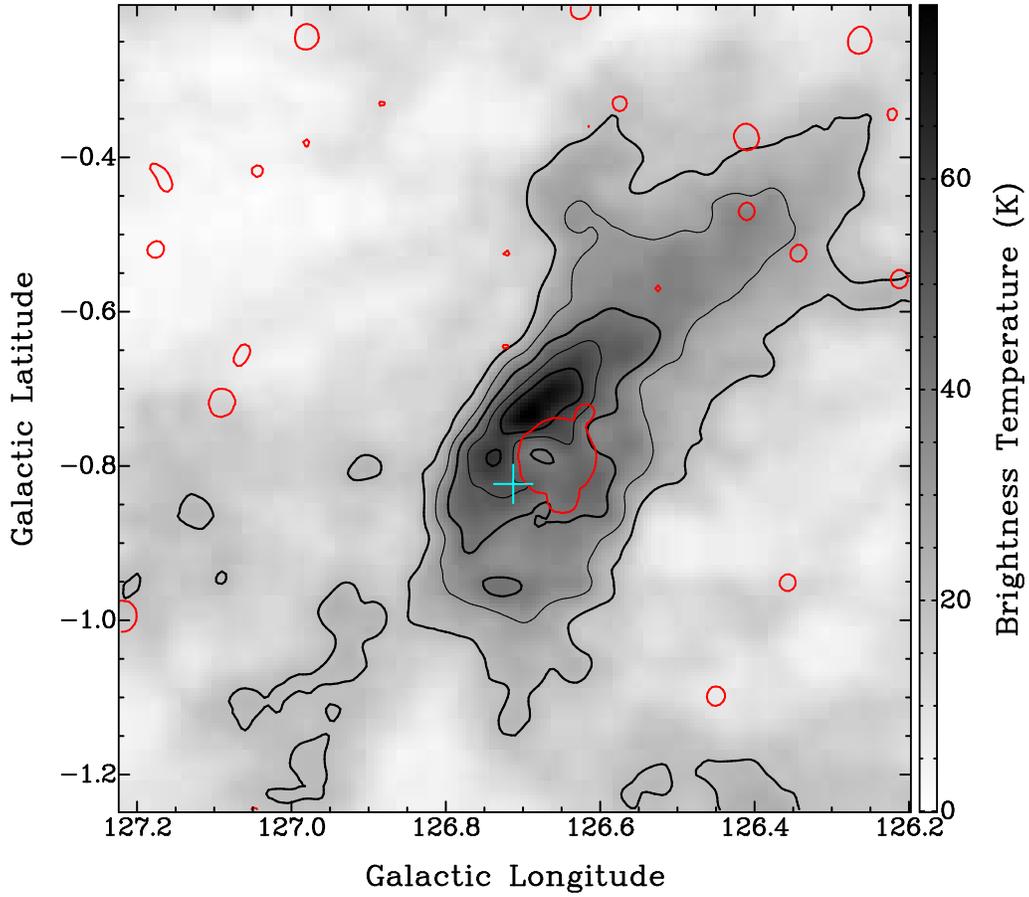}
\caption{The molecular cloud of KR 120. The image is $^{12}$CO ($J = 1\rightarrow 0$) emission from the
  Outer Galaxy Survey \citep{hey98} integrated between $-20.4 < V_{LSR} < -5.6$
  km~s$^{-1}$ (contours at 20, 30, 40, 50, and 60 ~K km s$^{-1}$). The 
  position of KR 120 is shown by the red contour at $T_{1420} = 8$~K,
  and the position of IRAS 01202+6133 is indicated by the cross. \label{fig:cocloud}}
\end{figure}

The morphology of the molecular cloud surrounding KR~120 (see
Figure~\ref{fig:cocloud}) is consistent with the alternative scenario of
triggered collapse of a pre-exisiting molecular core. The highest
column density gas (e.g., $T_B>50$~K km s$^{-1}$), where one would
expect to find molecular cloud cores, lies in a ridge to one side of
the \ion{H}{2} region. In this picture a shock front being driven
into this gas by the expanding \ion{H}{2} region has overrun and
compressed an pre-existing core initiating the star formation 
process \citep{ber89,tho04}. 

We cannot rule out a spontaneous star formation scenario; it
is possible that the YSO happened to form at the correct
distance from the exciting star of KR~120 that it now lies on the edge
of the expanding \ion{H}{2} region. On balance though, we prefer the
triggered scenario as it naturally explains the close
spatial association of the \ion{H}{2} region and the very young Class I YSO.

\section{Conclusions} \label{sec:conc}

1. Our analysis of the HCO$^{+}$ ($J = 3\rightarrow 2$; 267.6 GHz) and
H$^{13}$CO$^+$ ($J = 3\rightarrow 2$; 260.3 GHz) spectra of  the
submm/IR source IRAS~01202+6133 shows that the blue-asymmetric HCO$^{+}$
line profile is almost certainly due to infall motions.

2. We have modified the analytic radiative transfer models of \citet{mye96}
and \citet{dev05} to allow for fitting of submm/mm line profiles that
have both self-absorption features and optically thin wings. We
applied these models to our HCO+ submm spectrum of
IRAS~01202+6133 and conclude that it is a young Class I YSO
with a substantial envelope undergoing slow infall. 

3. Based on its young evolutionary state ($\tau_{YSO}<\tau_{HII}$), and
its proximity to the \ion{H}{2} region KR 120 (Sh 2-187),
we think that the formation of IRAS 01202+6133 was triggered by the 
expansion of the \ion{H}{2} region. However, due to young age of the \ion{H}{2} region 
a  "Collect and Collapse" scenario is ruled out.

\acknowledgements
We would like to thank Dr. Lewis B. G. Knee for his assistance with
the processing of the JCMT data.

{\it Facilities} \facility{JCMT}



\clearpage

\clearpage

\clearpage

\clearpage

\clearpage

\clearpage

\clearpage

\clearpage

\begin{deluxetable}{ccccccc} 
\tabletypesize{\scriptsize}
\tablecaption{Best Fit Free Parameters of 2L-O model \label{tbl:best2lo}}
\tablehead{
\colhead{$\tau_{0}$} & \colhead{$V_{LSR}$ (km~s$^{-1}$)} &  \colhead{$T_{K}$ (K)} & \colhead{$V_{in}$(km~s$^{-1}$)} &  \colhead{$\sigma_{NT}$(km~s$^{-1}$)}  & \colhead{$\tau_{out}$} & \colhead{$\sigma_{out}$(km~s$^{-1}$)}}
\startdata
$11 \pm 0.2$ & $-13.74\pm0.01$ & $9.5 \pm 0.15$ & $0.02 \pm 0.003$ & $0.4 \pm 0.01$ & $0.35 \pm 0.01$ & $2.3 \pm 0.03$
\enddata
\tablecomments{Background temperature is set at $T_{b} = 2.725$  K. Uncertainties are obtained by a Monte Carlo method with 1000 repetitions.}
\end{deluxetable}

\clearpage

\begin{deluxetable}{cccccccc} 
\tabletypesize{\scriptsize}
\tablecaption{Best Fit Free Parameters of Hill-O model \label{tbl:besthillo}}
\tablehead{
\colhead{$\tau_{c}$} &\colhead{$V_{LSR}$  (km~s$^{-1}$)} &\colhead{$V_{in}$  (km~s$^{-1}$)} &\colhead{$\sigma$(km~s$^{-1}$)}  &\colhead{$T_{p}$ (K)} &\colhead{$T_{0}$ (K)} &\colhead{$\tau_{out}$}  &\colhead{$\sigma_{out}$(km~s$^{-1}$)} }
\startdata
$6.5\pm 0.1$ & $-13.74 \pm 0.01$ & $0.07 \pm 0.012$ & $0.47 \pm 0.02$ & $7.7 \pm 0.13$ & $3.2 \pm 0.22$ & $0.32 \pm 0.03$ & $2 \pm 0.06$
\enddata
\tablecomments{Background temperature is set at $T_{b} = 2.725$  K. Uncertainties are obtained by a Monte Carlo method with 1000 repetitions.}
\end{deluxetable}

\end{document}